Highly Stable Glasses of *cis*-Decalin and *cis/trans*-Decalin Mixtures

*Katherine R. Whitaker, Daniel J. Scifo*[1]*, and M. D. Ediger*[*]

Department of Chemistry, University of Wisconsin – Madison, Madison, Wisconsin 53706, USA

*Mathias Ahrenberg and Christoph Schick*

Institute of Physics, University of Rostock, Rostock 18051, Germany

ABSTRACT: *In situ* AC nanocalorimetry was used to measure the reversing heat capacity of vapor-deposited glasses of decahydronaphthalene (decalin). Glasses with low heat capacity and high kinetic stability, as compared to the corresponding liquid-cooled glass, were prepared from *cis*-decalin and from several *cis/trans*-decalin mixtures. This is the first report of highly stable glass formation for molecular mixtures. The 50/50 *cis/trans*-decalin mixture is the highest fragility material reported to produce an ultrastable glass. The 50/50 mixture exhibited high kinetic stability, with an ~500 nm film deposited at 116 K (0.86 $T_{g,}$) displaying a transformation time equivalent to $10^{4.4}$ times the structural relaxation time of the supercooled liquid at the annealing temperature. *Cis*-decalin and the decalin mixture formed stable glasses that had heat capacities as much as 4.5% lower than the liquid-cooled glass.

[1] Current address: Eisenmann Corporation, Crystal Lake, IL, 60014, USA

[*] To whom correspondence should be addressed. Email: ediger@chem.wisc.edu. Phone: (608) 262-7273

## Introduction

Glasses are an interesting class of solids that can be formed from organic, inorganic, polymeric, colloidal, and metallic components.[1-2] While the irregular local packing of glasses is sometimes perceived as a drawback of these materials, it is precisely this feature that allows the properties of glasses to be tuned significantly by changing the composition. This tunability is responsible for much of the technological impact of glasses. Another important difference between glasses and crystals is that the properties of a glass can depend significantly upon its preparation history. For example, a glass formed by rapid cooling from the liquid will generally be less stable than a glass formed by slow cooling, and both of these glasses will typically be different from glasses formed by physical vapor deposition.

Glasses produced by physical vapor deposition can exhibit extraordinary properties in comparison with liquid-cooled glasses. Glasses with high kinetic stability and high density can be prepared[3-7] by depositing at low rates onto substrates at temperatures near 0.85 $T_g$; here $T_g$ is the conventional glass transition temperature. Such an enhancement of properties would be expected if a glass were prepared by very slow cooling from the liquid state or by aging below $T_g$ for long times[8-9]. The observed properties for vapor-deposited glasses though are equivalent to those expected for glasses aged for one thousand to one million years. These unique glasses also have low enthalpy[3, 5, 7], low heat capacity $C_p$,[7, 10-13] low water vapor uptake[14], high mechanical modulii[15-16], and can be anisotropic[17]. It is hypothesized that these materials can form because strongly enhanced mobility at the glass surface allows molecules to find efficient packing arrangements before they are buried by further deposition.[4] This mechanism is supported by theoretical work[18-19], by computer simulations[20-22], and by direct measurements of surface mobility[23].

The formation of highly stable glasses by physical vapor deposition appears to be quite general, with more than a dozen organic molecules reported to show this behavior to date. Vapor-deposited stable glasses have been formed from larger molecules including the family of tris-naphthylbenzene isomers[24] and the pharmaceuticals indomethacin and nifedipine[3, 5, 25], as well as smaller molecules such as toluene and ethylbenzene[7, 13, 26-28]. Much work has been done to investigate and characterize these unusually stable glasses in the last five years, but two avenues that remain unexplored are high fragility glassformers and mixtures. Both of these are investigated here for the first time.

Fragility is a topic that has received much attention in the field of supercooled liquids and glasses.[29] Fragility describes the temperature dependence of the viscosity or structural relaxation time as a material is cooled towards $T_g$. The kinetic fragility can be characterized by the steepness index m.[30] Glassformers that exhibit Arrhenius behavior have small values of m and are termed "strong"; glassformers that exhibit strongly non-Arrhenius behavior have large m values and are considered "fragile".[29, 31] Fragility provides a means of classifying glassformers and has been found to be correlated to the boson peak[32] and the shape of the potential energy landscape[33-34]. The relationship between fragility and the heat capacity jump at $T_g$ has also been extensively investigated.[35-37] Up to this point, all the systems shown to form stable glasses have intermediate fragility. Recently, it was reported that glycerol (the organic glassformer with the lowest fragility) does not form a stable glass when vapor-deposited near 0.85 $T_g$.[38] In two recent studies, water (which is often considered to be a low fragility liquid at low temperature) was vapor-deposited at different substrate temperatures and measured with fast scanning calorimetry[39-40]; stable glass features were observed in neither study. These results raise the question as to whether stable glass formation is possible across the range of known fragilities or if it might be restricted to a subset of fragilities.

Here we report experiments in which we vapor-deposit thin glassy films of *cis*-decalin and *cis/trans*-decalin mixtures over a range of substrate temperatures $T_{sub}$ and analyze the properties of these glasses with *in situ* AC nanocalorimetry. Kinetic stability is quantified by comparing the onset temperature $T_{onset}$ for the reversing heat capacity of the as-deposited glass to that of the liquid-cooled glass. For the 50/50 *cis/trans*-decalin mixture, the kinetic stability was further investigated by performing isothermal annealing experiments. Nanocalorimetry also enables us to measure the difference between the heat capacity of an as-deposited glass and the glass cooled from the liquid; $C_p$ for stable glasses can be significantly less than that for liquid-cooled glasses. These experiments on *cis*-decalin and *cis/trans*-decalin mixtures allow us to test whether stable glasses can be formed from molecular mixtures and also whether they can be formed from highly fragile glassformers. The 50/50 *cis/trans*-decalin mixture has the highest reported fragility among molecular glassformers.[41-42]

We show here that *cis/trans*-decalin mixtures with a wide range of compositions form stable glasses when vapor-deposited. The as-deposited glasses of the 50/50 mixture are up to 2.5% lower in heat capacity and have onset temperatures up to 7 K higher than that of the glass prepared by cooling from the liquid. The transformation time for a thin film (~500 nm) was $10^{4.4}$ times the structural relaxation time of the supercooled liquid at the annealing temperature. *Cis*-decalin forms stable glasses with up to 4.5% lower heat capacity for glasses deposited around 0.73 $T_{g,}$. Relative to the ordinary glass, the onset temperature of *cis*-decalin glasses can be increased 8 K by depositing between 0.84-0.94 $T_g$. Efforts to prepare a stable glass of *trans*-decalin were not successful, presumably due to crystallization during deposition.

This is the first report of a stable glass formed from a mixture and also establishes that very high fragility systems can form stable glasses. Stable glass mixtures could be technologically relevant in organic electronics[43-44] or amorphous pharmaceuticals[45-46].

**Experimental Methods**

**Materials.** *Cis*-decahydronaphthalene (*cis*-decalin) and *trans*-decahydronaphthalene (*trans*-decalin) were purchased from Sigma (St. Louis, MO) with a purity of 99%. Both were used without further purification. Figure 1 shows the molecular structures of the geometric isomers.

H
H
*trans*-decalin
H
H
*cis*-decalin

**Figure 1.** Structures of *trans*-decalin (left) and *cis*-decalin (right).

**Sample preparation.** *Cis*-decalin and *cis/trans*-decalin mixture films were prepared by physical vapor deposition. The liquid material was placed in crucibles maintained at room temperature outside of the vacuum chamber and a leak valve was used to control the flow of gaseous material into the chamber. The vapor pressure of the liquids is high enough that no heating of the crucibles was required to achieve the desired deposition rate.

Decalin and decalin mixtures were deposited directly from the vapor phase onto the nanocalorimeters (Xensor Integration, XEN-39391), which were held in a temperature-controlled housing. The temperature was measured with a surface resistance temperature detector (RdF Corporation) epoxied to the surface of the housing. Temperature was maintained by heating a cartridge heater against liquid nitrogen cooling. The temperature calibration is described below. The copper housing contained two nanocalorimeters; one calorimeter (the reference calorimeter) was completely enclosed by the housing while the other (the sample calorimeter) was located 2.8 mm below a 1.6 mm diameter opening that allowed deposition onto the active area. A mechanically rotatable shield located inside the vacuum chamber controlled deposition onto the sample calorimeter. For the measurements reported here, three different sample/reference calorimeter pairs were used. Similar results were achieved with all three pairs.

Prior to deposition, the pressure in the chamber was $10^{-10}$ torr. To deposit, the leak valve was opened until the ion gauge pressure (Granville-Phillips 274 Nude Bayard-Alpert Ionization Gauge) increased to 4.0 x $10^{-6}$ torr. [Note that this is the nitrogen equivalent pressure. Based on the deposited thickness, the actual decalin pressure is estimated to be approximately a factor of three lower than the gauge pressure.] After the pressure stabilized, the shield was rotated to allow deposition onto the sample sensor. Since deposition occurs through a small opening, the molecules can only approach the nanocalorimeter substrate at a cone of angles within 15° of normal; previous work has shown that the properties of vapor-deposited glasses can depend upon the angle at which the molecules approach the substrate.[47] The distance between the opening and the point where molecules enter the chamber through the leak valve is 20.5 cm. There is a line of sight from the leak valve to the calorimeter substrate and a portion of the molecules directly deposit in this manner. The lock-in amplifier signal (see below) was monitored during the deposition and once the desired increase in signal was observed, the shield was raised to end the deposition.

The absolute thickness of the film on the nanocalorimeter was determined by comparison with *in situ* ellipsometry measurements. For this comparison only, a completely exposed nanocalorimeter was placed adjacent to a silicon wafer. Simultaneous calorimetry and ellipsometry measurements were performed during several depositions. The relationship between the calorimetry signal and the thickness in these calibration experiments was then used to determine the film thickness for deposition through the 1.6 mm hole. For the results discussed in this paper, the *cis*-decalin and decalin mixture films were 570 ± 60 nm in thickness and deposited at a rate of 0.2 ± 0.1 nm/s.

The *cis/trans*-decalin mixture films were produced by co-deposition. *Cis*-decalin and *trans*-decalin were in separate crucibles, each with a leak valve. To deposit the 50/50 mixture, first the leak valve connected to the *cis*-decalin crucible was opened until the ion gauge pressure read $2.0 \times 10^{-6}$ torr. Then the leak valve of the *trans*-decalin crucible was opened until the total pressure came to the standard deposition pressure of $4.0 \times 10^{-6}$ torr; under these conditions, the gas phase composition in the chamber was 50% *cis*-decalin and 50% *trans*-decalin. Finally, the shield was positioned to allow deposition onto the nanocalorimeters. Since the deposition rate can be controlled within 3%, we assume the composition condensing on the calorimeters matches the intended composition within 3%, i.e., for the mixture above, (50 ± 3) % of the molecules on the nanocalorimeter are *cis*-decalin. Other compositions were achieved by changing the deposition pressures of each component accordingly.

***In situ* AC nanocalorimetry.** Our *in situ* AC nanocalorimetry measurements on decalin glasses closely follow those described previously.[48] Here we use the commercially available sensors XEN-39391 (Xensor Integration, The Netherlands) in an electrical set-up similar to references[49-50]. A digital lock-in amplifier (SR7265, Signal Recovery) measures the amplitude and phase of the complex differential thermopile signal. The complex amplitude of the differential thermopile signal is directly proportional to the reversing heat capacity of the sample. A differential set-up is used here for increased sensitivity and to minimize the influence of the addenda heat capacity.[51] To account for imbalances between the two sensors, an empty scan was performed prior to the measurements under the same conditions and subtracted in the complex plane from the sample measurement.

Nanocalorimetry measurements were performed during sample deposition and during subsequent temperature scanning. A thermal frequency of 20 Hz was used in the AC nanocalorimetry measurements to maximize sensitivity. In each experiment, the sample was deposited at the desired substrate temperature and then the temperature was changed at 2 K/min to 105 K. A 2 K/min heating ramp was then performed to obtain the heat capacity curve of the as-deposited glass (Figures 2, 3 and 4). Immediately after the dynamic glass transition was complete, the liquid was cooled at 2 K/min to form the liquid-cooled glass (also described herein as the “ordinary glass”). All the systems studied here were prone to crystallization so a rapid transition from heating to cooling was necessary and temperature scans could only be performed to a temperature just past the dynamic glass transition. Heating and cooling ramps at 2 K/min were then repeated a second time. Generally, the heat capacities observed during the second heating and the two cooling cycles were in excellent agreement; in these cases, the second heating curve was used as the reference curve for the liquid-cooled glass. In some experiments, particularly on pure *cis*-decalin samples, samples began to crystallize near the end of the second heating curve; in these cases, the first cooling curve was used as the reference for the liquid-cooled glass.

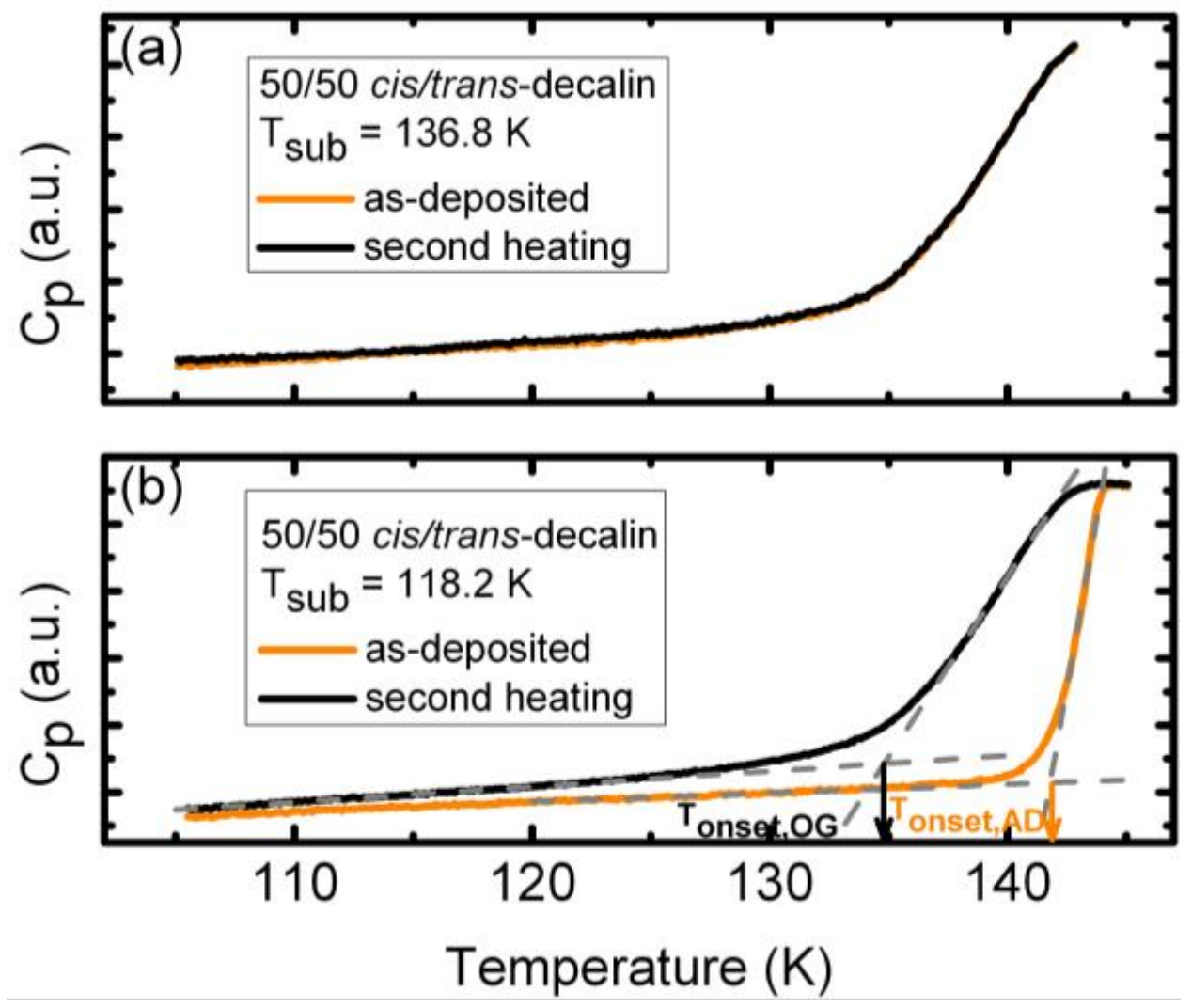

**Figure 2.** Effect of substrate temperature on the heat capacity of vapor-deposited 50/50 *cis/trans*-decalin mixture glasses. (a) Reversing heat capacity of the glass deposited at $T_{sub}$ = 136.8 K (1.02 $T_g$). The as-deposited glass (orange) has the same heat capacity as the glass prepared by cooling the liquid (black). (b) Reversing heat capacity of the glass deposited at $T_{sub}$ = 118.2 K (0.88 $T_g$). The as-deposited glass (orange) is lower in heat capacity and has a higher onset temperature than the glass cooled from the liquid (black). For both samples the scanning rate was 2 K/min and the thermal frequency was 20 Hz. The gray dotted lines indicate the tangent method used to determine the onset temperatures. The films were 570 ± 60 nm thick and were deposited at 0.2 ± 0.1 nm/s.

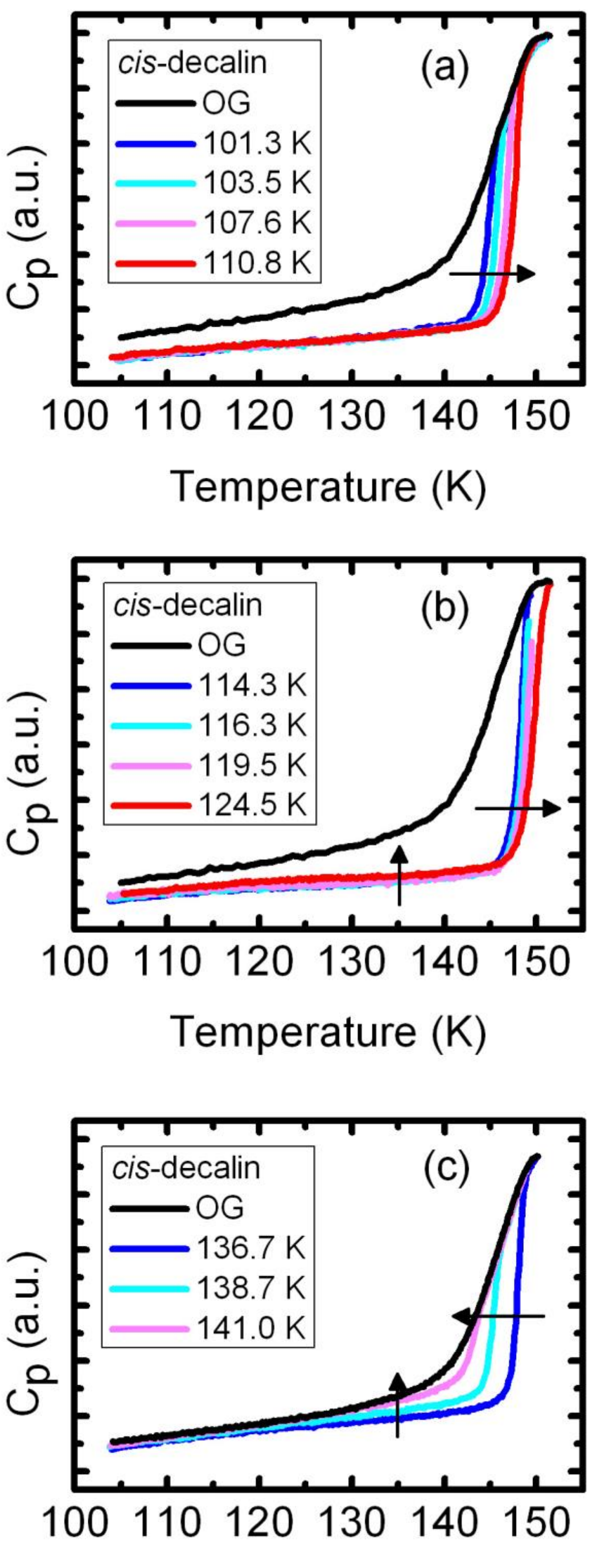

cis-decalin
(a)
OG
101.3 K
103.5 K
107.6 K
110.8 K
Cp (a.u.)
100 110 120 130 140 150
Temperature (K)
cis-decalin
(b)
OG
114.3 K
116.3 K
119.5 K
124.5 K
Cp (a.u.)
100 110 120 130 140 150
Temperature (K)
cis-decalin
(c)
OG
136.7 K
138.7 K
141.0 K
Cp (a.u.)
100 110 120 130 140 150
Temperature (K)

**Figure 3.** Effect of substrate temperature on the reversing heat capacity for vapor-deposited glasses of *cis*-decalin at substrate temperatures of (a) 0.72 to 0.79 $T_g$ (b) 0.81 to 0.88 $T_g$ and (c) 0.97 to 1.00 $T_g$. For comparison, the black line in each panel shows the heat capacity of the ordinary glass prepared by cooling the liquid. The arrows indicate the direction of increasing substrate temperature. For all samples the thermal frequency was 20 Hz and the scanning rate was 2 K/min. The *cis*-decalin films were deposited at 0.2 ± 0.1 nm/s at the substrate temperature specified in the legend and the thickness was 570 ± 60 nm.

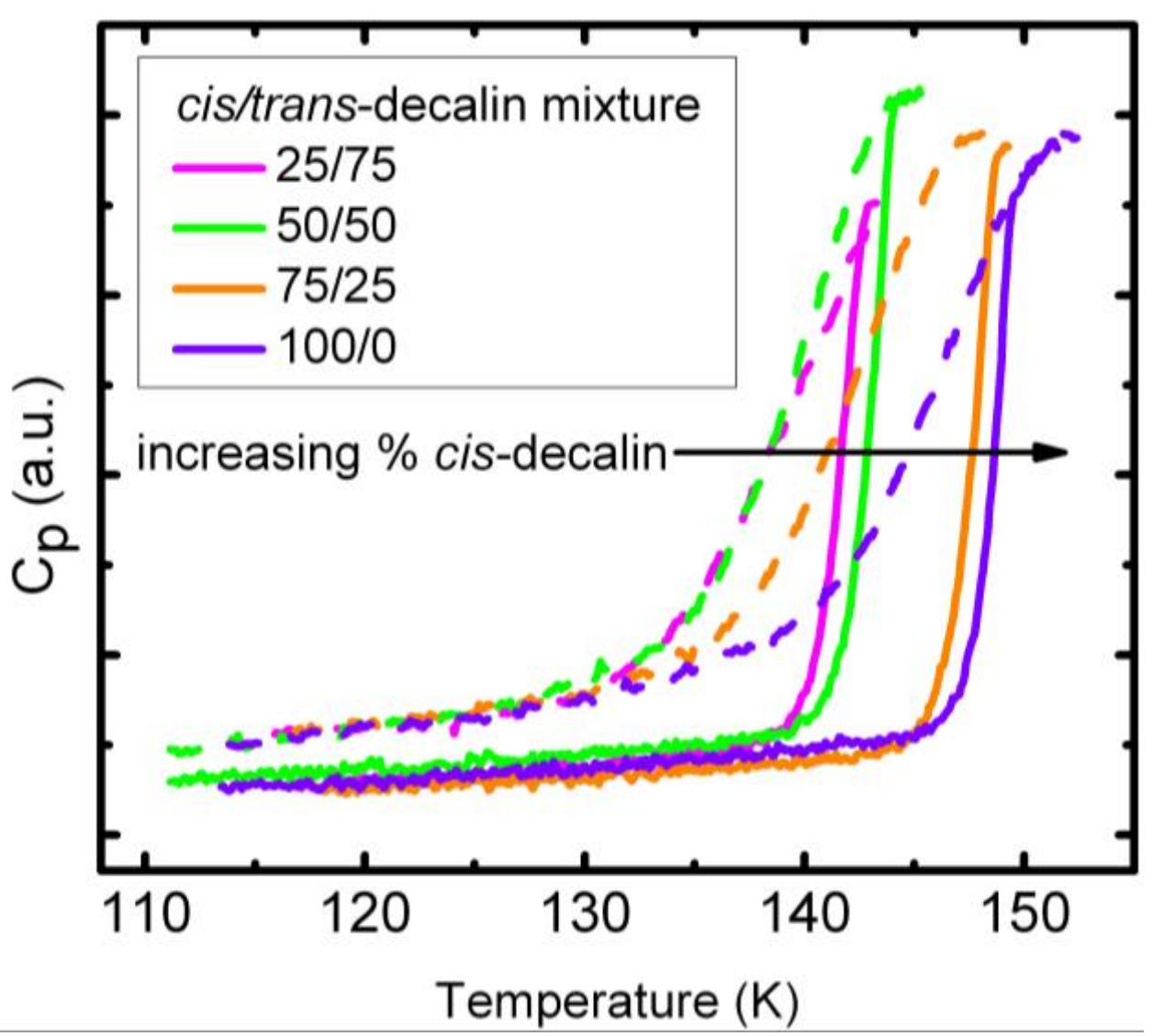

**Figure 4.** Effect of composition on the reversing heat capacity of *cis/trans*-decalin mixtures. Solid lines are heating curves of the as-deposited glasses and dotted lines are cooling curves of the corresponding liquid-cooled glass. The magenta curves are a 25/75 *cis/trans*-decalin mixture (deposited at 115.1 K), the green curves are a 50/50 mixture (deposited at 108.9 K), orange curves are a 75/25 mixture (deposited at 117.4 K) and the purple curves are pure *cis*-decalin (deposited at 113.0 K).

For the 50/50 *cis/trans*-decalin mixture, quasi-isothermal AC nanocalorimetry experiments were also performed. In these experiments, the sample was deposited at the desired substrate temperature and then the temperature was ramped at 5 K/min to the annealing temperature. The sample was held quasi-isothermally (~0.5 K temperature oscillation at 20 Hz) at the annealing temperature until the heat capacity of the sample was observed to stop changing. Subsequent temperature ramping experiments verified that this final state was the equilibrium supercooled liquid.

As a result of small thickness differences, the ordinary glass reference curves for different samples differ in amplitude by as much as 11%. To simplify the visual comparison of different samples in Figures 3, the ordinary glass curves were scaled to coincide at 119.5 K. The same small adjustments were also applied to the corresponding as-deposited glass $C_p$ values. Small vertical shifts were also applied to the data in Figure 4. The deposited film can cause small amounts of stress in the nanocalorimeter membrane, which in turn causes a change in the sample heater resistance. This effect was carefully measured and the reported heat capacities have been corrected for this.

**Temperature calibration.** We utilized the resistance of the heater on the nanocalorimeter membrane in order to determine the membrane temperature. Following ref [48], we calibrated the heater resistance using cyclopentane and toluene. Cyclopentane has three phase transitions in the relevant temperature range: 122.23 K (crystal—plastic crystal II), 138.35 K (plastic crystal II—plastic crystal I) and 178.59 K (plastic crystal I—liquid).[52] Ahrenberg et al.[48] have shown that the dynamic glass transition of toluene measured with AC nanocalorimetry is in good agreement with the dielectric relaxation data of Hatase et al.[53] and thus it can be used for temperature calibration.

As a check on the temperature calibration procedure, we make use of dielectric relaxation data for the supercooled 50/50 *cis/trans*-decalin mixture.[41] Using the temperature calibration described above, our nanocalorimetry measurements show that the 20 Hz dynamic glass transition temperature for the supercooled liquid of this mixture is on average 140.8 K, which is in good agreement with the value of 140.2 K reported in the dielectric work. The run-to-run variation of the dynamic glass transition of the supercooled liquid in our measurements is about 1 K. All ordinary glass reference curves were horizontally shifted to correct for these run-to-run variations and the same shift was applied to the corresponding as-deposited glass. For pure *cis*-decalin, $T_{g,dyn}(20\ Hz)$ was determined to be 147.5 K, with similar run-to-run variation.

A further check on the nanocalorimeter temperature calibration can be made by a comparison with differential scanning calorimetry (DSC) experiments. DSC experiments utilizing scanning rates of 1 K/min to 5 K/min were performed on *cis*-decalin and the 50/50 decalin mixture. The onset temperatures for the glass transition from these experiments were slightly extrapolated to obtain the 10 K/min values of 141.0 K and 134.8 K, respectively; these values are denoted as $T_g$ in this paper. The temperature differences between the glass transition temperatures measured with DSC and at 20 Hz with nanocalorimetry are in good agreement for the two systems, as would be expected if the fragility of *cis*-decalin is similar to that of the 50/50 mixture. In the DSC measurements, crystallization of the *cis*-decalin sample was avoided by fast-quenching the sample from room temperature; this was done by placing the room temperature DSC pan containing *cis*-decalin on the cold block of the DSC (maintained at 113 K). Similar DSC experiments were attempted for pure *trans*-decalin but were unsuccessful due to crystallization.

## Results

**Reversing $C_p$ during temperature ramping.** Vapor-deposited glasses can be equivalent to ordinary glasses or they can be highly stable, depending on the deposition temperature. This is illustrated for the 50/50 *cis/trans*-decalin mixture in Figure 2. Here we show the reversing heat capacity curves of decalin mixture glasses deposited at 1.02 $T_g$ and 0.88 $T_g$. In AC nanocalorimetry, the signal is proportional to the reversing heat capacity and thus non-reversing features such as the enthalpy overshoot are not observed. Figure 2(a) shows that the glass deposited near $T_g$ has the same heat capacity as the glass cooled from the liquid at 2 K/min; this can be seen by comparing the as-deposited glass (orange curve) and the ordinary glass (black curve). As seen in Figure 2(b), the *cis/trans*-decalin mixture glass deposited at 0.88 $T_g$ exhibits properties of a highly stable glass: low heat capacity and high onset temperature (high kinetic stability). The heat capacity of the as-deposited glass is 1.8% lower than the liquid-cooled glass and the onset temperature is 7 K higher. Upon heating, the as-deposited glass transforms into the supercooled liquid. The sample was subsequently cooled and re-heated; the second heating curve shows the properties of the liquid-cooled glass for comparison.

As shown in Figure 3, glasses of *cis*-decalin were deposited over a range of substrate temperatures from 101.3 K to 143.9 K, or 0.72 to 1.02 $T_g$. The properties of the vapor-deposited glasses are significantly affected by the substrate temperature. For clarity, Figure 3 has been divided into three panels each of which shows the behavior of the liquid-cooled glass for comparison. The arrows indicate the direction of increasing substrate temperature. Figure 3(a) shows glasses deposited from 0.72 to 0.79 $T_g$. In this range, the onset temperature of the as-deposited glass increases with substrate temperature. The heat capacity has the lowest values of all substrate temperatures but no trend is observed between the experiments shown within the error of the measurement. In the middle of the substrate temperature range (Figure 3(b), 0.81 to 0.88 $T_g$) the onset temperature continues to increase with substrate temperature and is maximized in this range. The heat capacity increases with substrate temperature. Figure 3(c) shows these same trends continue for glasses deposited from 0.97 to 1.00 $T_g$. *Cis*-decalin is not a good glassformer and these measurements are the first report of the glass transition of a completely amorphous sample. A similar series of experiments was performed on the 50/50 *cis/trans*-mixture. The primary data showed the same qualitative features as those shown in Figure 3 and a summary of the results is presented below.

Two other compositions of *cis/trans*-decalin were vapor-deposited and observed to form stable glasses. Figure 4 shows the as-deposited heating curves and the first cooling curves for 25/75, 50/50, 75/25 and 100/0 *cis/trans*-decalin vapor-deposited glasses. Each as-deposited glass has a lower heat capacity and a higher onset temperature than the corresponding liquid-cooled glass, indicating that a highly stable glass has been obtained. The glass transition temperature is seen to increase with increasing proportions of *cis*-decalin.

**Dependence of glassy $C_p$ and kinetic stability upon substrate temperature.** Vapor-deposited glasses of decalin can have lower heat capacity than the liquid-cooled glass. Figures 2 and 3 demonstrate this for the 50/50 *cis/trans*-decalin mixture and *cis*-decalin, respectively. In order to quantify this difference in heat capacity between the as-deposited and ordinary glasses, we use the quantity $C_{p,AD}/C_{p,OG} - 1$, determined at 0.85 $T_g$ (119.5 K and 114.5 K for *cis*-decalin and the 50/50 *cis/trans*-decalin mixture, respectively). The heat capacity difference is calculated at this temperature because the response here comes purely from the glassy solid (at the modulation frequency of 20 Hz). Figure 5 shows the fractional $C_p$ decrease as a function of substrate temperature during deposition. The substrate temperature is scaled by $T_g$ of the corresponding supercooled liquid. *Cis*-decalin and the decalin mixture show qualitatively similar trends. At substrate temperatures close to $T_g$ the heat capacities of the as-deposited glasses are almost equivalent to the corresponding liquid-cooled glasses, consistent with Figure 2(a). As the substrate temperature is decreased, the heat capacity of the as-deposited glass decreases. Near the lowest substrate temperatures used here, the fractional $C_p$ decrease is 4.5 ± 1% for *cis*-decalin and 2.5 ± 0.5% for the *cis/trans*-decalin mixture.

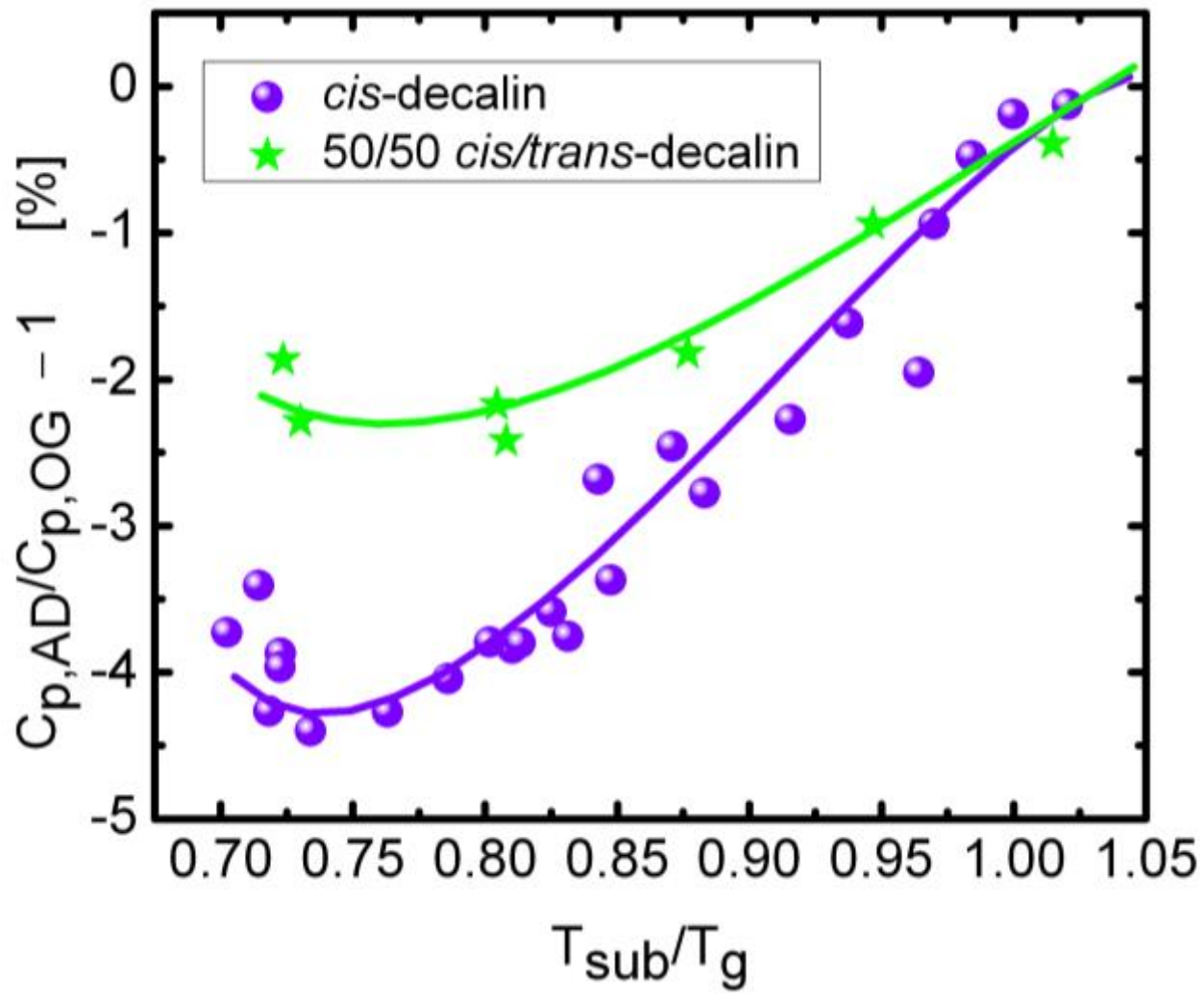

**Figure 5.** Heat capacity of as-deposited glasses (AD) of decalin, expressed as a percentage change relative to the ordinary glass (OG) prepared by cooling the liquid. The heat capacities used in the calculation were determined at 0.85 $T_g$ and the substrate temperature is normalized to $T_g$. The purple spheres are *cis*-decalin data and the green stars are 50/50 decalin mixture data. The solid lines are guides to the eye.

The onset temperature of the as-deposited glass also depends upon the substrate temperature. The onset temperature is determined according to the tangent method shown in Figure 2(b). Figure 6 compares the onset temperatures of the as-deposited glass to that of the liquid-cooled glass, as a function of substrate temperature. The trends in onset temperature are very similar for *cis*-decalin and the 50/50 *cis/trans*-decalin mixture. Close to $T_g$, the onset temperatures are equal to those of the ordinary glass. Between 0.84 and 0.94 $T_g$, the greatest increase of the onset temperature is seen and ranges up to 8 ± 1 K for *cis*-decalin and 7 ± 1 K for the decalin mixture. As the substrate temperature is further decreased, the onset temperature of the as-deposited glass begins to decrease.

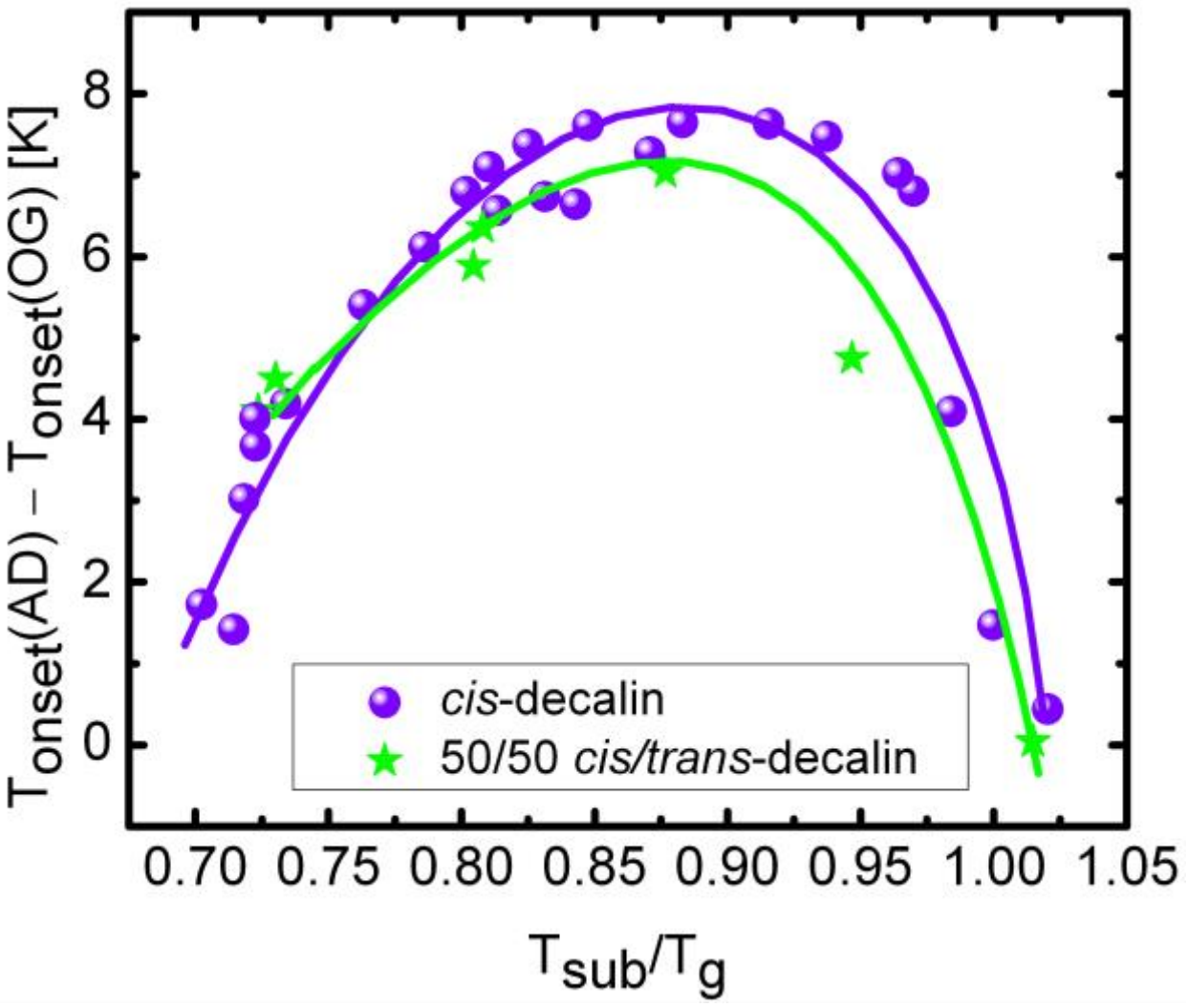

**Figure 6.** The onset temperatures for the as-deposited decalin glasses (AD) as a function of substrate temperature, relative to the ordinary glass (OG) cooled from the liquid. Purple spheres are *cis*-decalin data and green stars are decalin mixture data. The substrate temperature is normalized to $T_g$. Heating ramps were done at 2 K/min and the onset temperatures were determined by the tangent method, as shown in Figure 2(b). The solid lines are guides to the eye.

**Isothermal annealing experiments.** Isothermal annealing experiments were performed to determine the kinetic stability of vapor-deposited *cis/trans*-decalin mixture glasses. AC nanocalorimetry is ideally suited to study the isothermal transformation of a thin as-deposited glass to the supercooled liquid. A quasi-isothermal annealing experiment was carried out at 136.6 K on a 50/50 *cis/trans*-decalin mixture glass deposited at 0.86 $T_g$ (115.9 K). During annealing the heat capacity increases as the sample transforms from the glass to the supercooled liquid. $\Phi_{SG}(t)$ is used to represent the fraction of the sample responding as a stable glass at any point in time.[54] At the start of the annealing measurements $\Phi_{SG}(t = 0\ s) = 1$ and the sample is completely glassy. As the transformation proceeds, $\Phi_{SG}(t)$ decreases until the sample is entirely supercooled liquid and $\Phi_{SG}(t)=0$. The time when $\Phi_{SG}(t)$ reaches zero is defined as the transformation time of the as-deposited glass. The transformation shown in Figure 7 took ~43,000 s (determined by linear extrapolation), which is $10^{4.4}$ times the structural relaxation time $\tau_\alpha$ of the equilibrium supercooled liquid at the annealing temperature, as determined from dielectric spectroscopy.[41]

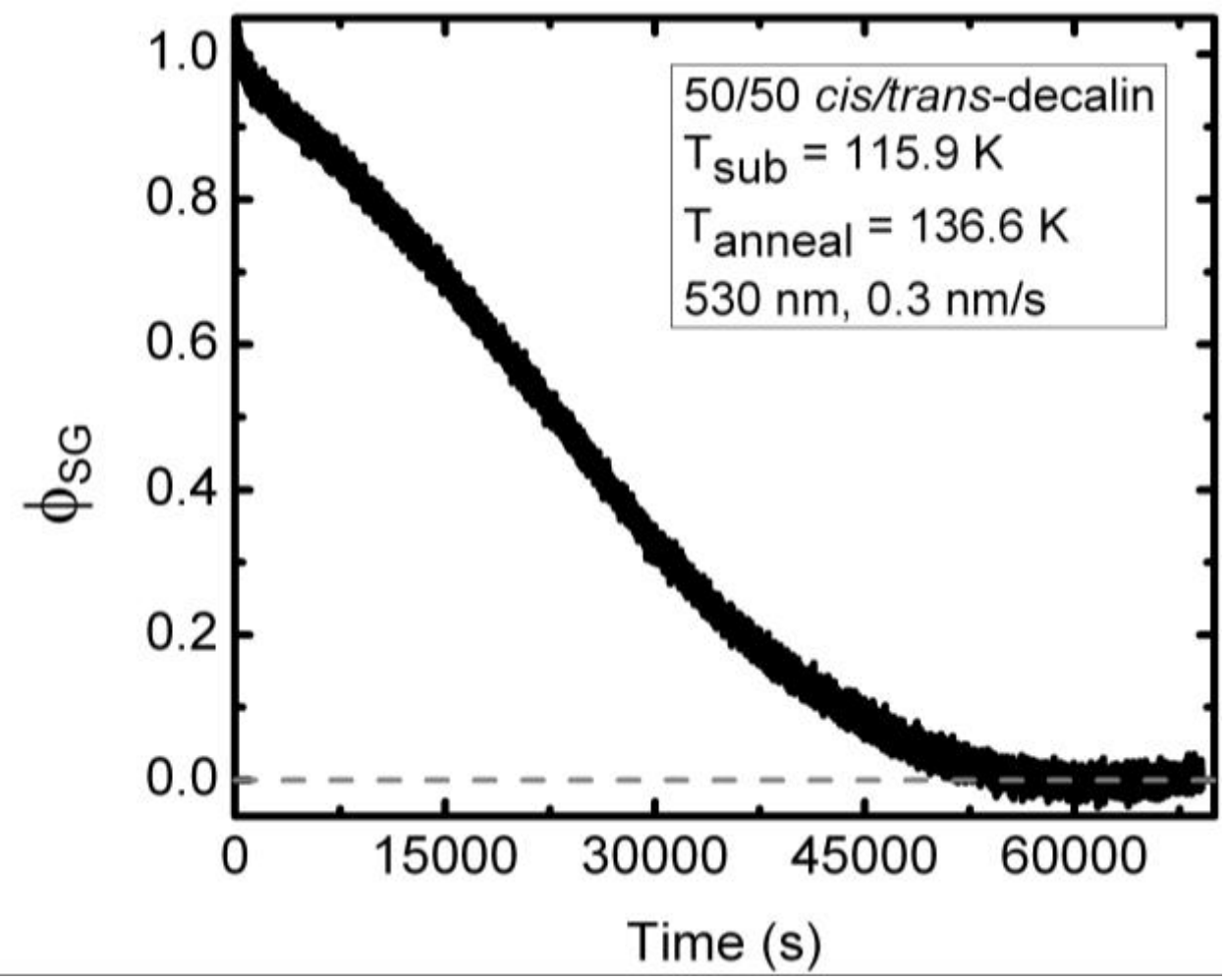

**Figure 7.** Isothermal annealing of a vapor-deposited 50/50 *cis/trans*-decalin mixture film (530 nm). $\Phi_{SG}$ represents the fraction of stable glass remaining at any point in time. This sample was deposited at 0.3 nm/s at 0.86 $T_g$ and annealed at 136.6 K.

## Discussion

**Substrate temperature dependence of vapor-deposited glasses.** During vapor deposition onto a substrate held near 0.85 $T_g$, molecules condensing onto the sample surface find a sufficiently mobile environment such that they have the opportunity to explore the energy landscape and find stable configurations. As a result of the deposition process, each molecule is part of the mobile surface layer before being buried in the film. These glasses become well-packed[4] from the bottom up.

The efficient packing that occurs during deposition near 0.85 $T_g$ gives rise to glasses that are kinetically stable and have low heat capacity. In order for thin films of these glasses to transform into the supercooled liquid, a growth front moves in from the free surface, providing the mobility needed for further transformation as it progresses.[19-20, 55] With nanocalorimetry, this growth front is most directly observed in isothermal experiments on thin films; a growth front results in a transformation curve that is approximately linear in time and there is a linear thickness dependence to the transformation time.[12, 56] Both of these features are consistent with the surface-initiated growth front mechanism observed with secondary ion mass spectrometry.[55] Highly stable vapor-deposited glasses have low heat capacity[7, 10-13] and it has been suggested that the higher packing efficiency of the stable glass shifts some of the vibrational modes to higher frequencies[12]. The resulting difference in the vibrational density of states (VDOS) would lead to the observed lower heat capacity of the stable glass. Simulations and experiments qualitatively support this idea. Mossa et al. showed for the Lewis-Wahnstrom model of ortho-terphenyl that the VDOS of higher density glasses was shifted to higher frequency.[57] Experimental work on a mineral glass showed that high fictive temperature glasses have VDOS shifted to lower frequency.[58] As stable glasses are characterized by higher density and lower fictive temperature relative to ordinary glasses, both of these observations are consistent with the VDOS of stable glasses being shifted to higher frequency and thus having a lower heat capacity.

The two preceding paragraphs provide the context needed to understand the properties of as-deposited decalin glasses as a function of substrate temperature. At substrate temperatures near $T_g$, the onset temperature is nearly the same as that of the ordinary glass (see Figure 6). Although there is adequate mobility at the glass surface to allow rapid configurational sampling[23], the temperature is sufficiently high that equilibration does not result in highly efficient packing. As the substrate temperature is lowered, the glasses become more kinetically stable until the substrate becomes so cold that the molecules lack the mobility to sample different packing arrangements. For both *cis*-decalin and the *cis/trans*-decalin mixture, maximum kinetic stability is observed 0.84-0.94 $T_g$. These results are similar to what Ahrenberg et al.[13] observed with AC nanocalorimetry for ethylbenzene and toluene and those reported by Kearns et al. [3] in DSC experiments on indomethacin.

The heat capacity of the vapor-deposited glasses follows a somewhat different trend than the onset temperature over the range of substrate temperatures studied (see Figure 5). Similar to the onset temperature, the heat capacity is equal to that of the liquid-quenched glass at substrate temperatures closest to $T_g$, but below this region, the heat capacity decreases significantly (Figure 5). Near the lowest substrate temperatures, the heat capacity has a maximum decrease of ~2.5% and ~4.5% for the *cis/trans*-decalin mixture and *cis*-decalin, respectively. We expect that if we were able to access even lower deposition temperatures, we would see the heat capacity increase again, as was observed for toluene and ethylbenzene.[13] The data in Figure 5 suggests an increase in heat capacity would occur just beyond the lower end of our accessible temperature range. At very low substrate temperatures, molecules do not have time to find more stable configurations so the heat capacity should increase. Consistent with this view, Suga and coworkers[59-62] showed that the enthalpy of a glass vapor-deposited at very low temperatures is higher than the enthalpy of the liquid-cooled glass.

For vapor-deposited toluene and ethylbenzene glasses, Ahrenberg et al.[13] examined the relationship between the heat capacity and the onset temperature over a range of substrate temperatures. For both systems, the two quantities were anti-correlated down to ~0.80 $T_g$, with increasing onset temperatures associated with decreasing heat capacity. At lower substrate temperatures, this anti-correlation is lost as both the onset temperature and the heat capacity decrease. These qualitative observations for toluene and ethylbenzene are consistent with the results presented here for *cis*-decalin and the 50/50 *cis/trans*-decalin mixture. This behavior is not understood and merits further exploration. In this context, we note that decalin is the first molecule without an aromatic ring in its structure to be shown to form stable glasses.

**Stable glasses and fragility.** Table 1 provides a compilation of some of the properties of vapor-deposited stable glasses, including results from the literature and those reported here. In order to compare the kinetic stability of different materials, we tabulate $t_{trans}$ for a 500 nm stable glass film; this is the time required for the stable glass to transform into the supercooled liquid at a temperature where the structural relaxation time of the supercooled liquid $\tau_\alpha$ is about 1.5 s. $t_{trans}$ is expressed as a ratio relative to $\tau_\alpha$; to a first approximation, this ratio expresses the factor by which the stability of the stable glass exceeds that of an ordinary liquid-cooled glass. Table 1 also shows the heat capacity decrease for the stable glass in comparison with the ordinary liquid-cooled glass. For each of these quantities, results are shown for the deposition conditions that maximize the tabulated property. For one case, ααβ-tris-naphthylbenzene, the value of $t_{trans}/\tau_\alpha$ shown in the table was obtained by interpolation of the published data.

**TABLE 1: Vapor-deposited glasses measured with AC nanocalorimetry**

| system | $T_g$ (K) | fragility[a] | % $C_p$ decrease[b] | $t_{trans}/\tau_\alpha$ [c] | references |
|---|---|---|---|---|---|
| indomethacin | 315 | 83 | 4.5 ± 2 | $10^{4.1}$ | 5, 10, 63-65 |
| ααβ-tris-naphthylbenzene | 348 | 86 | 4 ± 1 | $10^{4.4}$ | 5, 12, 63, 65 |
| toluene | 117 | 104 | 4 ± 0.5 | $10^{3.8}$ | 13, 63, 66 |
| ethylbenzene | 115 | 97 | 4 ± 0.5 | $10^{3.7}$ | 13, 63, 66 |
| *cis*-decalin | 141 | - | 4.5 ± 1 | - | this work |
| 50/50 *cis/trans*-decalin | 135, 137 | 145, 147 | 2.5 ± 0.5 | $10^{4.4}$ | 41-42, this work |

[a] The fragility column reports $m = \left.\frac{d\log\tau}{d(T_g/T)}\right|_{T=T_g}$

[b] $(1 - C_{p,AD}/C_{p,OG}) \times 100\%$
[c] $t_{trans}$ is the transformation time for a 500 nm film at the temperature where $\tau_\alpha \approx 1.5$ s for the supercooled liquid.

Stable glass formers cover a range of fragilities, as can be seen in Table 1. The 50/50 *cis/trans*-decalin mixture has the highest fragility of molecular glassformers. As determined by calorimetry and dielectric relaxation spectroscopy, the values of "m" characterizing the kinetic fragility are 145[42] and 147[41], respectively. Since surface mobility is a key factor in the formation of stable glasses, we can infer from Table 1 that organic glasses show substantial surface mobility across a wide range of fragilities, including very fragile systems.

**Stable glasses of mixtures.** These are the first experiments to show that stable glasses can be formed from mixtures. Figure 4 shows vapor-deposited glasses of decalin mixtures over a range of possible compositions. For each composition, the as-deposited glass has a lower heat capacity and a higher onset temperature than the corresponding liquid-cooled glass, demonstrating that a stable glass is formed. The ability to continuously tune the properties of glasses by changing composition is one of the key features that make glasses technologically useful. An important element of the current work is the demonstration that stable glasses can also show this versatility. If the properties that we associate with stable glasses, such as low heat capacity[7, 10-13], high density[4] and an extra peak in the wide-angle X-ray scattering[67-68] were somehow a result of nanocrystals rather than a truly amorphous system, this new result would be difficult to understand. While mixed molecular crystals can be formed in some cases, they are uncommon especially if the molecules are not superimposable[69] and a mixed crystal of *cis*- and *trans*-decalin has not been reported. Thus Figure 4 strongly argues against the idea that nanocrystals are responsible for the extraordinary features of vapor-deposited glasses. Additional arguments also support this view.[68]

The mixed stable glasses of *cis*- and *trans*-decalin have properties that are distinct from the pure components. Stable glasses of *cis*-decalin and the 50/50 *cis/trans*-decalin mixture have different heat capacity decreases at low substrate temperature, with *cis*-decalin achieving a larger heat capacity decrease (see discussion below). Additionally, Table 1 shows that among the systems investigated, the mixed decalin stable glass has the smallest heat capacity decrease, relative to the ordinary glass. While it was not possible to form glasses from vapor-deposited *trans*-decalin, stable glasses could be formed that contained up to 75% *trans*-decalin in the mixture. This illustrates the possibility to incorporate the molecules of a very poor glassformer into stable glasses at quite high concentration. This opportunity might be relevant in organic electronics[44] or for fast-crystallizing drug molecules.[45-46]

Can any mixture form a stable glass? The mixture of two geometric isomers of decalin is quite simple; the two molecules have the same chemical functionality and the glass transition temperatures differ by only about 5%. It will be interesting to see if stable glasses can also be formed from molecules that differ more significantly, e.g., in terms of glass transition temperatures. Perhaps for molecules of quite different sizes, stable glass formation might still be possible in a limited composition window. A key component of stable glass formation is molecular mobility during deposition. When the glass transition temperatures are different enough, it may not be possible to find a substrate temperature where both components have sufficient mobility to support stable glass formation. It may also be interesting to investigate whether the ability to form a stable glass of a given mixture will be hindered or facilitated by the degree of intermolecular attractions between the components.

**Stability of *cis*-decalin vs. decalin mixture glasses.** The kinetic stability, as determined by onset temperature, is quite similar for *cis*-decalin and 50/50 *cis*/*trans*-decalin glasses but the heat capacity decrease is markedly different. As seen in Figure 6 both *cis*-decalin and the decalin mixture show a maximum onset temperature increase of 7-8 K at substrate temperatures around 0.88 $T_g$. Moreover, across the entire substrate temperature range the onset temperatures and thus kinetic stability are comparable. The fractional $C_p$ decrease of *cis*-decalin and the 50/50 mixture, however, begin to deviate at substrate temperatures below 0.95 $T_g$ (Figure 5). At the lowest substrate temperatures, the *cis*-decalin glasses show almost twice the decrease in heat capacity as compared to the ordinary liquid-cooled glass as the decalin mixture. Considering that the heat capacity and onset temperature do not appear to be correlated for a single substance as deposition conditions are varied, it is reasonable that the kinetic stability of the two systems can be similar while the fractional $C_p$ decreases differ. It is interesting that the decalin mixture has a smaller decrease in heat capacity than any of the materials shown in Table 1. We anticipate that investigations of additional systems will reveal which factors control the observed heat capacity drop. For reference, we note that physical aging has previously been shown to lower $C_p$ for glasses, but aging on the order of days only lowers $C_p$ by ~ 1% or less.[72-75]

**The generality of stable glass formation.** Stable glass formation is not limited to good glassformers. *Cis*-decalin is a very poor glassformer and for this reason few reports on glasses or supercooled liquids of the pure component can be found in the literature. Prior to the DSC results that we report here, the glass transition temperature was only measured on a sample that was more than 95% crystalline.[76] Our measurements provide the first reported glass transition temperature of pure *cis*-decalin for a completely amorphous sample. *Cis*-decalin joins βββ-TNB[24] in the ranks of poor glassformers that can be vapor-deposited to form stable glasses. However, we also attempted to vapor-deposit glasses of pure *trans*-decalin, which based on our lack of success in DSC measurements, is an even poorer glassformer. A *trans*-decalin film could be formed on the calorimeter by vapor-deposition but the signal changes that occurred upon heating were not consistent with transformation into the supercooled liquid; we interpret this result to mean that the as-deposited film was already largely crystalline or crystallizes in the vicinity of $T_g$.

Crystallization during deposition represents one way in which stable glass formation can fail. A recent paper by Capponi et al.[38] describes another route by which stable glass formation can fail. These authors vapor-deposited glycerol onto substrates near 0.88 $T_g$ at a rate of 0.03 nm/s. Based upon previous work, as summarized in Table 1, stable glass formation would be expected under these conditions. In contrast, Capponi et al. did not obtain a stable glass. When they heated their as-deposited glass above $T_g$ it transformed relatively quickly into a liquid with unusually large dipolar order; this liquid persisted for an extremely long time. Qualitatively similar behavior was observed by these authors when they vapor-deposited threitol and xylitol. In light of the range of fragility of these molecules, we are inclined to interpret their behavior in terms of the dense network of hydrogen bonds formed by these liquids. In view of this, the failure of water to form stable glasses[39-40] could also be seen as a result of its hydrogen bonding network. Why liquids with a strong network of hydrogen bonds might behave systematically different than the substances listed in Table 1 is a matter for future investigation. One possibility is that these liquids do not have highly mobile surfaces. In any case, it remains an open question as to whether any moderately strong organic glassformer can form a highly stable glass via vapor-deposition.

**Conclusions**

*In situ* nanocalorimetry has been used to measure the heat capacity of glasses of *cis*-decalin and *cis/trans*-decalin mixtures vapor-deposited across a range of substrate temperatures. For all systems investigated, glasses deposited very near $T_g$ had properties identical to those of the corresponding liquid-cooled glasses. For both *cis*-decalin and 50/50 *cis/trans*-decalin mixtures, the heat capacity of the as-deposited glass decreased with decreasing substrate temperature. The largest differences were seen near the lowest deposition temperatures (~0.75 $T_g$), and were equal to a 4.5 ± 1% decrease for *cis*-decalin and a 2.5 ± 0.5% decrease for the decalin mixture. These vapor-deposited glasses exhibited increased kinetic stability as evidenced by increases in the onset temperature of up to almost 8 K relative to the liquid-cooled glass. Glasses deposited between 0.84 and 0.94 $T_g$ had the highest onset temperatures. Extremely long isothermal transformation times ($10^{4.4}$ $\tau_\alpha$ for ~500 nm film) were observed when 50/50 *cis/trans*-decalin glasses were annealed above $T_g$.

These results extend our knowledge of stable glass formation in at least two important directions. We have shown that extremely fragile glassformers can form stable glasses. Combined with previous results, stable glass formation has now been demonstrated for systems with intermediate and high fragility. Our results also show that mixtures of organic molecules can form stable glasses and that extremely poor glassformers can form stable glasses if mixed with ~25% of a second component. The ability to tune properties by changing composition is an important characteristic of glasses that enables many technological applications. Our results show that stable glasses also have this flexibility with regard to composition and this may facilitate applications of stable glasses. Amorphous mixtures are found, for example, in organic electronics and in pharmaceuticals in the form of drug-excipient mixtures that stabilize the drug. Stable glasses could potentially be useful in these applications and elsewhere that mixtures play a role. For fundamental studies, dyes or other probes could be incorporated into stable glasses to track the dynamics in the system. In addition, the formation of stable glasses from mixtures implies that nanocrystalline order likely plays no role in the enhanced stability observed in these vapor-deposited glasses.

**Acknowledgement.** We would like to thank Gilbert Nathanson and Ranko Richert for helpful discussions. We would like to thank Prof. Richert in particular for suggesting the implications of these results for the hypothesis of nanocrystalline order. We thank Shakeel Dalal for performing the *in situ* ellipsometry measurements and Yeong Zen Chua for support regarding temperature calibration. We gratefully acknowledge funding from the US National Science Foundation (CHE-0724062 and CHE-1012124) and the German Science Foundation (DFG-SCHI 331 14-1).